\documentclass[conference]{IEEEtran}
\usepackage{amsmath,subfigure}
\usepackage{graphicx}
\usepackage{grffile}
\usepackage{algpseudocode}
\usepackage{algorithm}
\usepackage{epstopdf}

\newcommand{\qed}{\nobreak \ifvmode \relax \else
      \ifdim\lastskip<1.5em \hskip-\lastskip
      \hskip1.5em plus0em minus0.5em \fi \nobreak
      \vrule height0.75em width0.5em depth0.25em\fi}

\begin{document}
\title{Multiple Hypotheses Iterative Decoding of LDPC in the Presence of Strong Phase Noise}

\author{
\IEEEauthorblockN{Shachar Shayovitz and Dan Raphaeli} \\
\vspace{-0.3cm}
\IEEEauthorblockA{Dept. of EE-Systems, Tel-Aviv University \\
Tel-Aviv 69978, Israel\\
Tel.\ 03-6408011 Fax 03-6405027  \\
Email: shachars@post.tau.ac.il,danr@eng.tau.ac.il}}

\vspace{-0.3cm}
\maketitle

Many satellite communication systems operating today employ low cost upconverters or downconverters which create phase noise. This noise can severely limit the information rate of the system and pose a serious challenge for the detection systems. Moreover, simple solutions for phase noise tracking such as PLL either require low phase noise or otherwise require many pilot symbols which reduce the effective data rate.

In the last decade we have witnessed a significant amount of research done on joint estimation and decoding of phase noise and coded information. These algorithms are based on the factor graph representation of the joint posterior distribution. The framework proposed in \cite{worthen2001}, allows the design of efficient message passing algorithms which incorporate both the code graph and the channel graph. The use of LDPC or Turbo decoders, as part of iterative message passing schemes, allows the receiver to operate in low SNR regions while requiring less pilot symbols.

In this paper we propose a multiple hypotheses algorithm for joint detection and estimation of coded information in a strong phase noise channel. We also present a low complexity mixture reduction procedure which maintains very good accuracy for the belief propagation messages.

\section{System Model}
We consider the transmission of a sequence of complex modulation symbols $\mathbf{c} = (c_{0},c_{1},...,c_{K-1})$
over an AWGN channel affected by carrier phase noise. We assume the symbols are drawn independency from an MPSK constellation. The discrete-time baseband complex equivalent channel model at the receiver is given by:
\begin{equation}\label{sys_model}
    r_{k} = c_{k}e^{j\theta_{k}}+n_{k} \;\;\;\;  k=0,1,...,K-1.
\end{equation}
The phase noise stochastic model is a wiener process:
\begin{equation}\label{weiner}
    \theta_{k} = \theta_{k-1} + \Delta_{k}
\end{equation}
where ${\Delta_{k}}$ is a real, i.i.d gaussian sequence with $\Delta_{k} \sim \textsl{N}(0,\sigma_{\Delta}^{2})$.

\begin{figure}
  \centering
  \includegraphics[width=8.5cm]{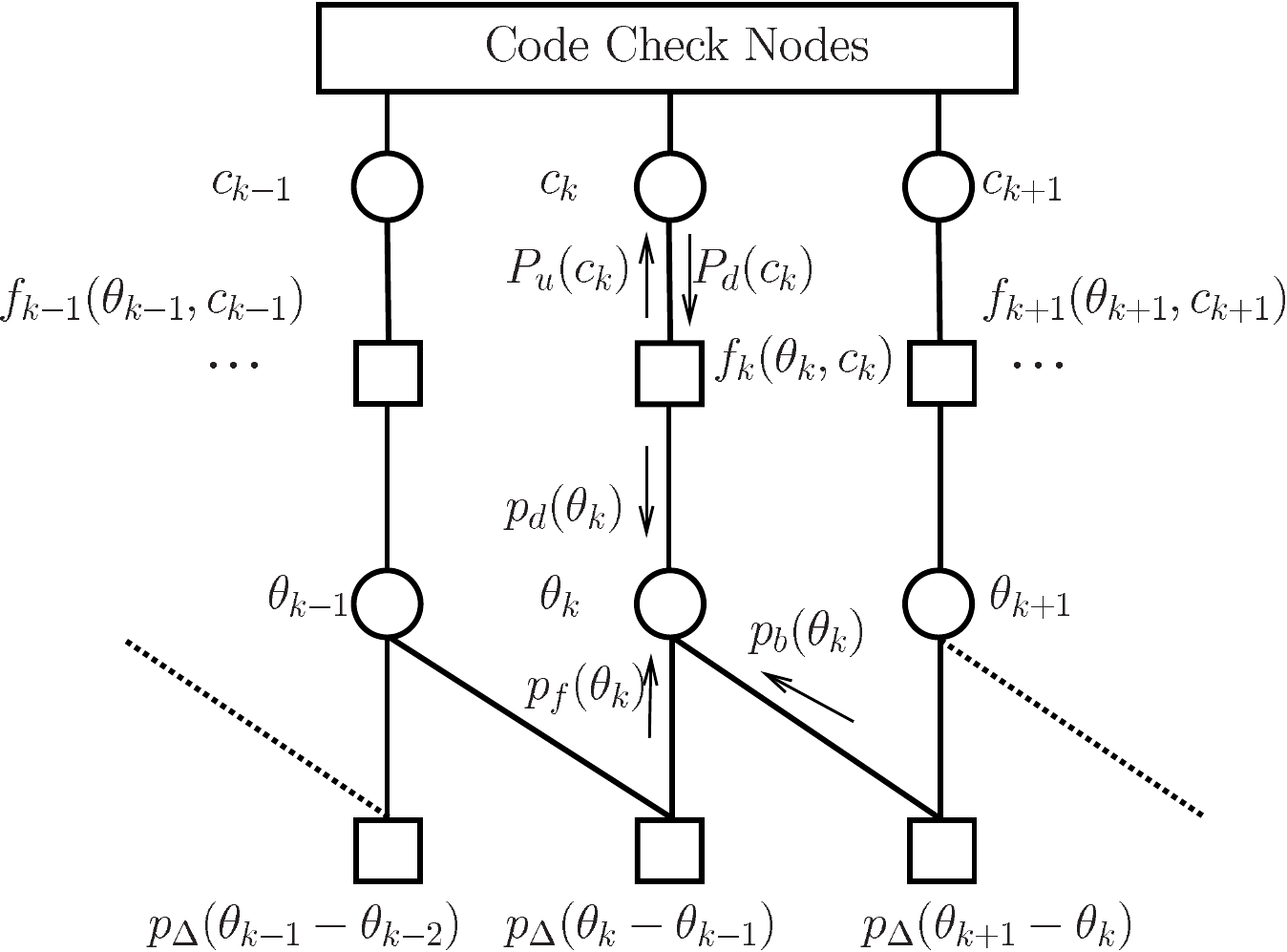}\\
  \caption{Factor graph representation of the joint posterior distribution}\label{fig:fg}
\end{figure}

The factor graph representation of the joint posterior distribution was given in \cite{barb2005} and is shown in Fig. \ref{fig:fg}.

The resulting Sum \& Product messages are:
\begin{equation}\label{pf}
    p_{f}(\theta_{k}) \propto \int_{0}^{2\pi}p_{f}(\theta_{k-1})p_{d}(\theta_{k-1})p_{\Delta}(\theta_{k}-\theta_{k-1})d\theta_{k-1}
\end{equation}
\begin{equation}\label{pb}
    p_{b}(\theta_{k}) \propto \int_{0}^{2\pi}p_{b}(\theta_{k+1})p_{d}(\theta_{k+1})p_{\Delta}(\theta_{k+1}-\theta_{k})d\theta_{k+1}
\end{equation}
\begin{equation}\label{pd}
    p_{d}(\theta_{k}) \propto \sum_{m=0}^{M-1} P_{d}(c_{k}=e^{j\frac{2\pi m}{M}}) f_{k}(c_{k},\theta_{k})
\end{equation}
\begin{equation}\label{Pu}
    P_{u}(c_{k}) \propto \int_{0}^{2\pi}p_{f}(\theta_{k})p_{b}(\theta_{k})f_{k}(c_{k},\theta_{k})d\theta_{k}
\end{equation}

\begin{equation}\label{fk}
    f_{k}(c_{k},\theta_{k}) \propto \exp\{-\frac{|r_{k}-c_{k}e^{j\theta_{k}}|^{2}}{2\sigma^{2}}\}
\end{equation}

\begin{equation}\label{p_del}
    p_{\Delta}(\theta_{k}) = \sum^{\infty}_{l=-\infty}g(0,\sigma_{\Delta}^{2},\theta_{k}-l2\pi)
\end{equation}

Where $M$,$r_{k}$,$P_{d}$, $\sigma^{2}$ and $g(0,\sigma_{\Delta}^{2},\theta)$ are the constellation order, received base band signal, symbol soft information from LDPC decoder, AWGN variance and Gaussian distribution respectively.

Due to the fact that the phase symbols are continuous random variables, a direct implementation of these equations is not possible and approximations are unavoidable. In \cite{shachar2012}, a modified Tikhonov approximation is used for the messages in the SPA which leads to a very simple and fast algorithm.

The best known message passing algorithm for phase noise channels quantizes the phase noise and performs an approximation of the sum \& product algorithm (SPA). This algorithm (called DP - discrete phase in this paper) requires large computational resources to reach high accuracy, rendering it not practical for some real world applications. In this paper, an approximate inference algorithm is proposed which better balances the tradeoff between accuracy and complexity for strong phase noise channels.

\section{Multiple Hypotheses Canonical Model}
We propose to approximate the SPA messages using the following Tikhonov mixtures,

\begin{equation}\label{new_pf}
    p_{f}(\theta_{k-1}) = \sum_{i =1}^{N}\alpha^{f}_{i}t^{f}_{i}(\theta_{k-1})
\end{equation}
\begin{equation}\label{new_pb}
    p_{b}(\theta_{k+1}) = \sum_{i =1}^{N}\alpha^{b}_{i}t^{b}_{i}(\theta_{k+1})
\end{equation}

Where:
\begin{equation}\label{f_f_i}
t^{f}_{i}(\theta) =  \frac{e^{Re[z^{f}_{i} e^{-j\theta}]}}{2\pi I_{0}(|z^{f}_{i}|)}
\end{equation}
\begin{equation}\label{f_b_i}
t^{b}_{i}(\theta) =  \frac{e^{Re[z^{b}_{i} e^{-j\theta}]}}{2\pi I_{0}(|z^{b}_{i}|)}
\end{equation}

And,
$\alpha^{f}_{i}$ ,$\alpha^{b}_{i}$,$z^{f}_{i}$,$z^{b}_{i}$ are the mixture coefficients and Tikhonov parameters of the forward and backward messages of the phase symbol $\theta_{k-1}$ and $\theta_{k+1}$ respectively. $N$ is the model order for both messages.

If we insert approximations (\ref{new_pf}) and (\ref{new_pb}) in to the forward and backward recursion equations (\ref{pf}), (\ref{pb}) respectively, we get

\begin{equation}\label{pf_eq}
    p_{f}(\theta_{k}) =
    \int_{0}^{2\pi}(\sum_{i =1}^{N}\alpha^{f}_{i}t^{f}_{i}(\theta_{k-1}))p_{d}(\theta_{k-1})p_{\Delta}(\theta_{k}-\theta_{k-1})d\theta_{k-1}
\end{equation}
\begin{equation}\label{pb_eq}
    p_{b}(\theta_{k}) = \int_{0}^{2\pi}(\sum_{i =1}^{N}\alpha^{b}_{i}t^{b}_{i}(\theta_{k+1}))p_{d}(\theta_{k+1})p_{\Delta}(\theta_{k+1}-\theta_{k})d\theta_{k+1}
\end{equation}

The resulting Tikhonov mixtures will be of order $NM$. Therefore, a mixture reduction algorithm must be derived which captures ''most" of the information in the mixtures (\ref{pf_eq}) and (\ref{pb_eq}), while keeping the computational complexity low.

We define the following mixture reduction task using the Kullback Leibler divergence - Given a Tikhonov mixture $f(\theta)$ of order $L$, find a Tikhonov mixture $g(\theta)$ of order $N$ ($L>N$), which minimizes,

\begin{equation}\label{obj}
D_{KL}(f(\theta)||g(\theta))
\end{equation}

Where,

\begin{equation}\label{KL_mix}
D_{KL}(f(\theta)||g(\theta)) = \int_{0}^{2\pi}f(\theta)\log{\frac{f(\theta)}{g(\theta)}}d\theta
\end{equation}

\begin{equation}\label{orig_mix}
f(\theta)=  \sum_{i =1}^{L}\alpha_{i}f_{i}(\theta)
\end{equation}
\begin{equation}\label{red_mix}
g(\theta)=  \sum_{j =1}^{N}\beta_{j}g_{j}(\theta)
\end{equation}

This divergence detailed in \cite{KL1951}, is a very popular measure of "closeness" of the reduced mixture to the original mixture. Unfortunately, the task of finding the mixture, $g(\theta)$  which minimizes (\ref{KL_mix}) is NP hard.
There are many suboptimal mixture reduction algorithms which work on the principle of sequentially merging mixture components until the target mixture size is reached. A very good summary of many of these algorithm can be found in \cite{sv2011}.

\subsection{Canonical Model - Adaptive Mixture Size}
Instead of reducing the mixture to a \textbf{fixed} size $N$, we propose a new approach which has better accuracy while keeping low complexity. Since we are performing bayesian inference on a large data block, we have many mixture reductions to perform (one for each symbol), rather than just a single reduction. Therefore, in terms of computational complexity, it is useful to use different mixture sizes for different symbols and look at the average number of components as a measure of complexity. We have shown that this new observation is critical in achieving high accuracy and low PER.

\subsection{Mixture Reduction Algorithm}
In this section, a mixture reduction algorithm will be proposed which receives as input
\begin{equation}\label{algo_input}
f(\theta)=  \sum_{i =1}^{L}\alpha_{i}f_{i}(\theta)
\end{equation}

Where,
$f_{i}(\theta)$ are Tikhonov distributions.

And outputs a reduced order Tikhonov mixture,

\begin{equation}\label{algo_output}
g(\theta)=  \sum_{j =1}^{N}\beta_{j}g_{j}(\theta)
\end{equation}

Where,
$g_{i}(\theta)$ are Tikhonov distributions and $N<L$.

The details of the algorithm are given in pseudo-code in Algorithm 1. This algorithm uses the Circular Mean and Variance Matching (CMVM) approach, detailed in \cite{shachar2012}, for optimally merging a Tikhonov mixture to a single Tikhonov distribution. Moreover, the function $|f(\theta)|$ outputs the number of Tikhonov components in the Tikhonov mixture $f(\theta)$.

\begin{algorithm}
\caption{Mixture Reduction Algorithm}
\label{mix_redc_algo}
\begin{algorithmic}
\State $j \gets 1$
\While{$j \leq L$  or  $|f(\theta)| > 0$}
\State $lead \gets argmax_{k}\{alpha_{k}\}$
\State $idx \gets lead$
 \For{$i = 1 \to |f(\theta)|$}
     \If {$D_{KL}(f_{i}(\theta) || f_{lead}(\theta)) \leq \mu $}
         \State $idx \gets [idx , i]$
    \EndIf
\EndFor
\State $g_{j}(\theta) \gets CMVM(\alpha(idx),f(idx))$
\State $\beta_{j} \gets \sum{\alpha(idx)}$
\State $f(\theta) \gets f(\theta) - \sum_{i \in idx}{\alpha(i)f_{i}(\theta)} $
\State Normalize $f(\theta)$
\State $j \gets j + 1$
\EndWhile
\end{algorithmic}
\end{algorithm}

The output Tikhonov mixture $g(\theta)$ is a reduced version of $f(\theta)$ and approximates the next forward and backward messages.

The choice of the threshold $\mu$ in the algorithm, is according to the level of distortion allowed for the reduced mixture with respect to the original mixture. If $\mu$ is very close to zero, then there won't be any components close enough and the mixture will not be reduced. Therefore, there is a tradeoff between complexity and accuracy in the selection of this parameter. Moreover, this parameter is not sensitive to SNR, phase noise variance or other scenario parameters which make this algorithm very robust.

We define $N$ (maximum allowable mixture order) to be very large ($ \approx 20$). In the simulations performed in this paper, the average number of components was much lower than the maximum allowed.

\subsection{Average Number of Mixture Components}
Using monte carlo simulations, we show that the average number of components in the canonical model, denoted in this paper as $\gamma$ is reasonably small compared to the constellation order. This size depends mainly on the number of ambiguities the phase estimation algorithm suffers between pilots. These ambiguities are a function of the SNR, phase noise variance and algorithmic design parameters such as the number of LDPC iteration and the threshold $\mu$.

The proposed algorithm shows, very small average number of mixture components per block symbol, thus keeping a reasonable computational complexity load.

\subsection{Complexity}
The complexity is summarized in table 1, where the computational load for DP and the algorithm proposed in this paper are compared. The computational complexity for DP was taken from \cite{barb2005}.
We assume that all non-linear operations are implemented using look up tables. The complexity analysis was done assuming that the number of components in the canonical model is the average number of mixture components per iteration $i$ - $\gamma(i)$, which is measured using simulations. It is important to note that $\gamma(i)$, the average number of hypotheses in iteration i, decreases as the iterations increases. This is due to the fact that the LDPC decoder provides better soft information on the symbols thus resolving ambiguities and decreasing the required number of components in the mixture.

\begin{table}[h]
\caption{Computational load per code symbol per iteration for M-PSK constellation}  

\centering  
\begin{tabular}{cccc}
\\[1ex] \hline\hline                       
  &DP & Multi Hyp Algo
\\ [0.5ex]
\hline\hline             
Operations & $13ML+10QL-9L-3M$ & $M\gamma(i)(11+5\gamma(i))+M$\\[1ex]
LUT & $3ML+2QL-3L-M$ & $M\gamma(i)(6+\gamma(i))$\\[1ex]
\hline                          
\end{tabular}
\label{tab:PPer}
\end{table}

Let $L$ be the number of quantization levels and $Q$ is a parameter for the DP algorithm explained in \cite{barb2005}.

We can see that we can achieve a high level of accuracy while maintaining a low computational load. Therefore, the algorithm proposed in this paper provides an improved tradeoff between accuracy and complexity.

\section{Numerical Results}
\begin{figure}
  \centering
  \includegraphics[width=8.5cm]{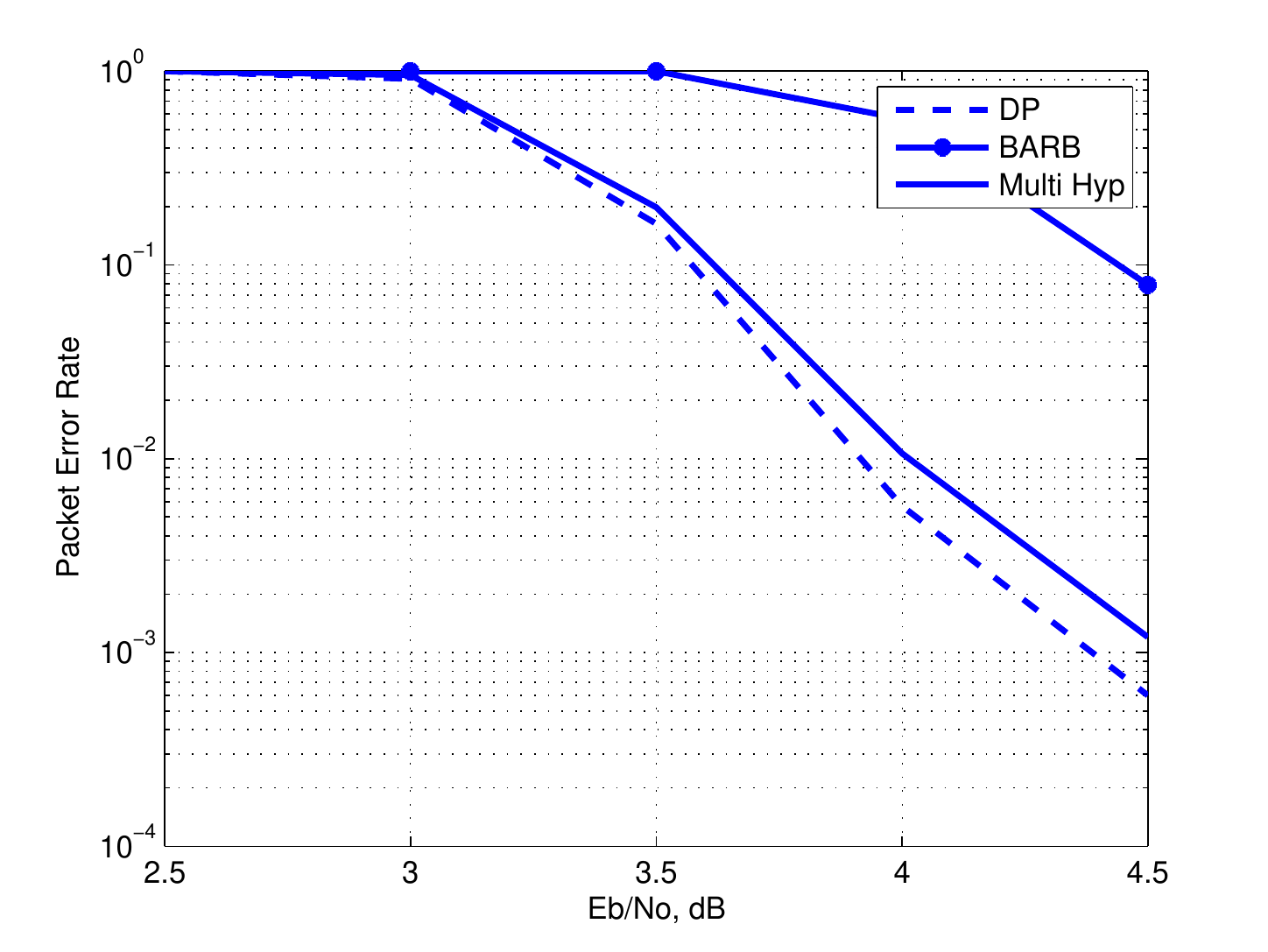}\\
  \caption{Packet error rate - QPSK and wiener phase noise}\label{fig:res}
\end{figure}

Monte Carlo simulation results for the algorithm are shown in Fig. \ref{fig:res}. A length 4608 LDPC code with rate 0.75 was used, where the coded bits were mapped to an QPSK constellation. The phase noise model used was a wiener process with $\sigma_{\Delta}=0.1$[rads/symbol]. A single pilot was inserted every 60 symbols. It should be noted that without pilots, the proposed algorithm can work as well as DP. The DP algorithm was simulated using 8 quantization levels. The parameter $\mu$ was adjusted empirically to achieve a good accuracy and complexity tradeoff.

As shown in Fig \ref{fig:res}, the proposed algorithm provides very good packet error rate in high phase noise level and very close to the performance of the optimal algorithm even when very few pilots are present and the code rate is high. The algorithm proposed in \cite{barb2005}, denoted BARB in the figure was added for comparison. It can be shown that the proposed multiple hypotheses algorithm performs much better than BARB, which is considered state of the art.

\begin{figure}
  \centering
  \includegraphics[width=8.5cm]{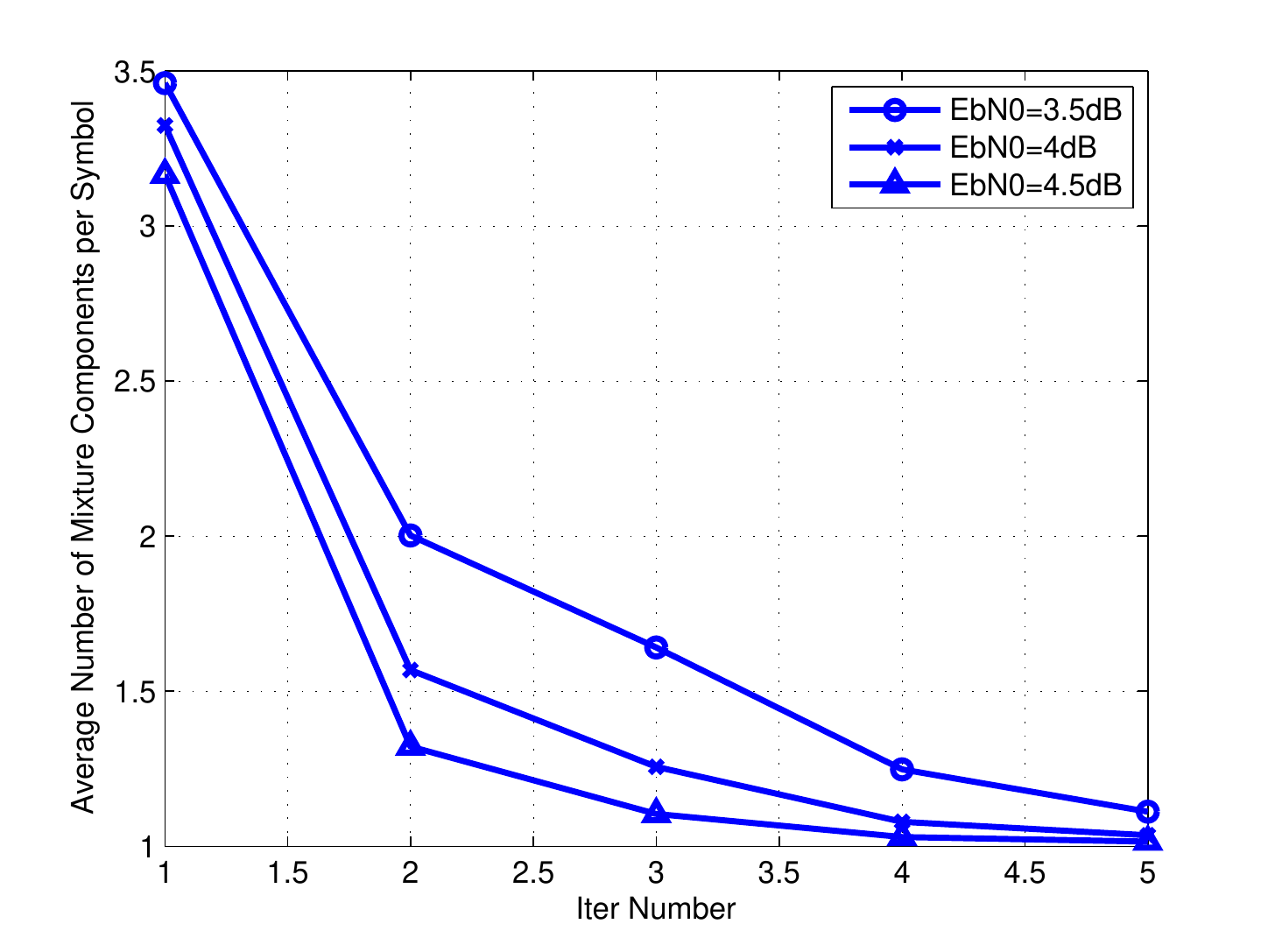}\\
  \caption{Average Mixture Order per Symbol}\label{fig:avg_comps}
\end{figure}

\begin{figure}
  \centering
 \includegraphics[width=8.5cm]{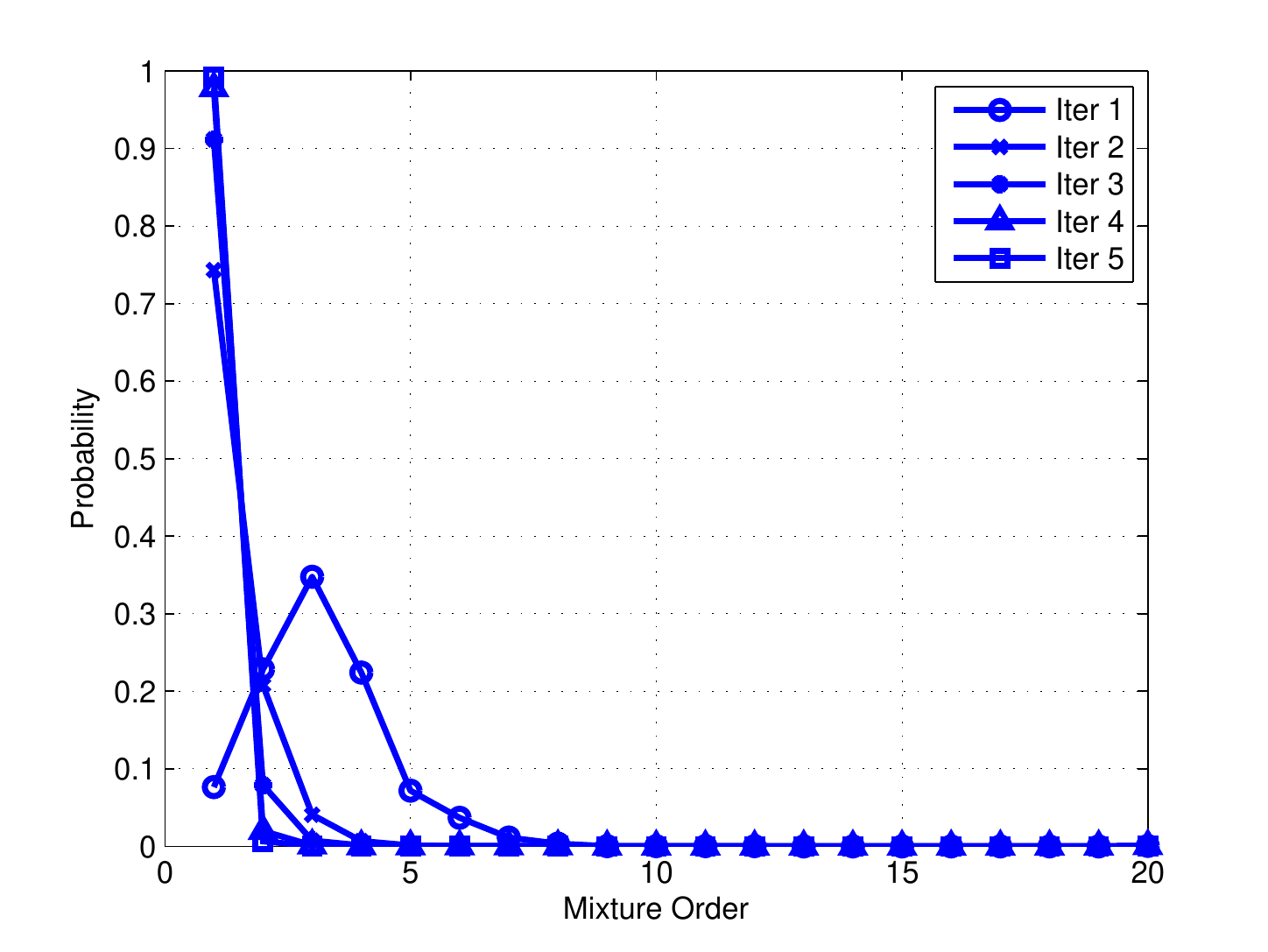}\\
 \caption{Distribution of Mixture Order - SNR $= 4.5dB$}\label{fig:res4_5db}
\end{figure}

Moreover, in Fig. \ref{fig:avg_comps}, we present the average number of mixture components in the canonical model for different SNR and LDPC iterations. It can be seen that for the first iteration, there is a need to have many components because there is a high level of phase ambiguity and in order to maintain accuracy the mixture order is greater than 1 ($3.5$ components per received symbol). As the iterations progress the LDPC decoder sends better soft information for the code symbols, resolving these ambiguities. Therefore, the average number of mixture components becomes closer to $1$.
In Fig. \ref{fig:res4_5db}, we show the distribution of the mixture order in the canonical model for different iterations for SNR $=4.5$dB. In this SNR, the packet error rate of the algorithm is around $10^{-3}$, so the distribution of the number of components is of real operational interest, since it creates an upper bound on the physical memory required to store the components. It can be seen, for this scenario, that the mixture order never exceeds 7 components.

In table 2, we show the computational complexity of the proposed algorithm in comparison to the DP algorithm. It should be noted that the computational complexity of the multiple hypotheses algorithm varies between LDPC iterations, thus reducing the overall complexity, while the computational complexity of the DP algorithm remains constant.

\begin{table}[h]
\caption{Simulation Results - Computational load per code symbol per iteration for M-PSK constellation}  

\centering  
\begin{tabular}{cccc}
\\[1ex] \hline\hline                       
  &DP & Multi Hyp, iteration 1 &Multi Hyp, iteration 2
\\ [0.5ex]
\hline\hline             
Operations & $2324$ & $403$ & $115$\\[1ex]
LUT & $476$ & $133$ & $45$\\[1ex]
\hline                          
\end{tabular}
\label{tab:PPer}
\end{table}

\section{Conclusions}
In this paper an algorithm for joint detection and estimation of LDPC in strong phase noise channels was presented. The proposed algorithm is based on the approximation of SPA messages using Tikhonov mixture canonical models. The problem of exponential increase in mixture size was solved using a new approach for mixture dimension reduction proposed in this contribution. This approach significantly reduced the computational complexity while keeping PER levels very close to the optimal algorithm (DP). The dimension reduction also uses newly reported results in directional statistics for optimal clustering of Tikhonov mixtures.




\bibliographystyle{IEEEtrans}
\bibliography{strings}

\end{document}